\documentclass[pdflatex,aps,prl,twocolumn,superscriptaddress,fleqn,showpacs]{revtex4-1}
\usepackage{times,xspace}
\usepackage{amsbsy,amssymb,amsmath,bm}
\usepackage{graphicx,color,epsfig,rotate}
\usepackage[usenames,dvipsnames]{xcolor}
\usepackage{longtable}
\usepackage{rotating}
\usepackage{soul}

\sethlcolor{Apricot}
 
\usepackage{fancyhdr}

\def\bbbc{{\mathchoice {\setbox0=\hbox{$\displaystyle\rm C$}\hbox{\hbox
to0pt{\kern0.4\wd0\vrule height0.9\ht0\hss}\box0}}
{\setbox0=\hbox{$\textstyle\rm C$}\hbox{\hbox
to0pt{\kern0.4\wd0\vrule height0.9\ht0\hss}\box0}}
{\setbox0=\hbox{$\scriptstyle\rm C$}\hbox{\hbox
to0pt{\kern0.4\wd0\vrule height0.9\ht0\hss}\box0}}
{\setbox0=\hbox{$\scriptscriptstyle\rm C$}\hbox{\hbox
to0pt{\kern0.4\wd0\vrule height0.9\ht0\hss}\box0}}}}

\newcommand{\bybal}{{$\beta$-YbAlB$_4$}}
\newcommand{\blbal}{{$\beta$-LuAlB$_4$}}
\newcommand{\aybal}{{$\alpha$-YbAlB$_4$}}

\newcommand{\yrs}{{YbRh$_2$Si$_2$}}

\newcommand{\rhall}{\rho_{xy}}

\newcommand{\Weiss}{$\Theta_{\rm W}~$}
 
\begin{document}

\title{Evolution of $c$-$f$ hybridization and two component Hall effect in \bybal}
 
\author{E. C. T. O'Farrell}
\email{eoin@issp.u-tokyo.ac.jp}
\affiliation{Institute for Solid State Physics, University of Tokyo, Kashiwa, Japan 277-8581}
\author{Y. Matsumoto}
\affiliation{Institute for Solid State Physics, University of Tokyo, Kashiwa, Japan 277-8581}
\author{S. Nakatsuji}
\email{satoru@issp.u-tokyo.ac.jp}
\affiliation{Institute for Solid State Physics, University of Tokyo, Kashiwa, Japan 277-8581}
 
\date{\today}
\begin{abstract}

\bybal~is the unique heavy fermion superconductor that exhibits unconventional quantum criticality without tuning in 
a strongly intermediate valence state. Despite the large coherence temperature, set by the peak of the longitudinal resistivity, 
our Hall effect measurements reveal that resonant skew scattering from incoherent local moments persists down to at least $\sim40$ K, where the Hall coefficient exhibits
a distinct minimum signaling another formation of coherence. The observation strongly suggests that the hybridization between $f$-moments and conduction
electrons has a two component character with distinct Kondo or coherence scales $T_K$ of $\sim40$ K and 200 K; this is confirmed by the magnetic field dependence of $\rhall$.
 
\end{abstract}
 
 
\maketitle
 

The Hall effect has a two-fold application in correlated electron systems; firstly, to extract material properties, and secondly, to
investigate electron-electron interactions and magnetism. In particular for $4f$ electron based intermetallics, the Hall effect has proved significantly useful to probe
how localized $4f$ moments in a metal host acquire a band character on cooling, through the Kondo effect.

Empirically, the total Hall resistivity $(\rhall)$ and Hall coefficient ($R_H$) are the sum of the normal and anomalous Hall effects (NHE and AHE) \cite{nagaosa2010anomalous},
\begin{align}
\label{eqn:1}
\rhall(T,H)&=R_{N}(T,H) H + R_{a}(T,H) M(T,H)\\
\label{eqn:2}
R_H&=R_N+R_a\chi\equiv R_N+R_A
\end{align}
where $T$, $H$,  $M$ and $\chi$ are temperature, applied magnetic field, magnetization, and magnetic susceptibility, respectively. 

While the NHE 
has a well understood dependence on the carrier density, Fermi surface (FS) and momentum dependent lifetime $\tau(\vec{k})$ \cite{nagaosa2010anomalous,ong1991geometric}, 
the AHE is a topic of recent intense research, which has revealed its topological nature \cite{nagaosa2010anomalous,fang2003anomalous,machida2009time}. 
Particularly, in mixed valence and Kondo systems, resonant skew-scattering is expected to dominate the AHE \cite{coleman1985theory,fert1987theory,kontani1994theory}.
 
In this Letter, we present Hall effect measurements of the quantum critical (QC) superconductor \bybal. 
Experimentally the material has striking properties, it exhibits QC without tuning i.e. at zero magnetic field and
ambient pressure \cite{nakatsuji2008superconductivity,Matsumoto2011Science};
the criticality is unconventional, and is not accounted for by the standard theory based on the spin-density wave description.
Moreover, it is the first example of superconductivity (with $T_c=80$~mK) among Yb based heavy fermion (HF) compounds \cite{nakatsuji2008superconductivity,kuga2008superconducting}. 
In sharp contrast with other known QC materials that have nearly integer valence, \bybal~has strong valence fluctuations with Yb$^{2.75+}$
and a characteristic high Kondo temperature $T_K = 250$ K \cite{nakatsuji2008superconductivity,okawa2010strong}. 

Normally, intermediate valence materials are magnetically inert at low temperatures reflecting a high Kondo temperature. 
In exception to this, \bybal~exhibits not only unconventional QC, but also Kondo lattice behavior characterized by a small renormalized scale of $T^* = 8$ K.
Here, as a key to resolve the dichotomy, Hall resistivity measurements find that incoherent skew scattering persists down to at least 40 K, one order of
magnitude smaller than $T_K = 250$ K. This signals the formation of another coherence due to the unquenched part of local $f$-moment.
Analyzing both $H$ and $T$ dependence of the $\rhall$ reveals two components of $f$-electrons that are characterized by two distinct Kondo
scales, $\sim 40$ and 250 K. We argue that the low energy component, which has low carrier density and incoherent behavior down to 40 K, is
responsible for both Kondo lattice and QC behaviors below $T^*$.


We performed AC transport using standard techniques. All measurements were made on high quality single crystals of \bybal with $H$ applied along the $c$ axis and current in the $ab$-plane.
Methods and analysis are described in more detail in the supplementary material.
 
 
\begin{figure}
\begin{center}
\includegraphics[viewport=100 15 478 427,clip=true,width=\columnwidth]{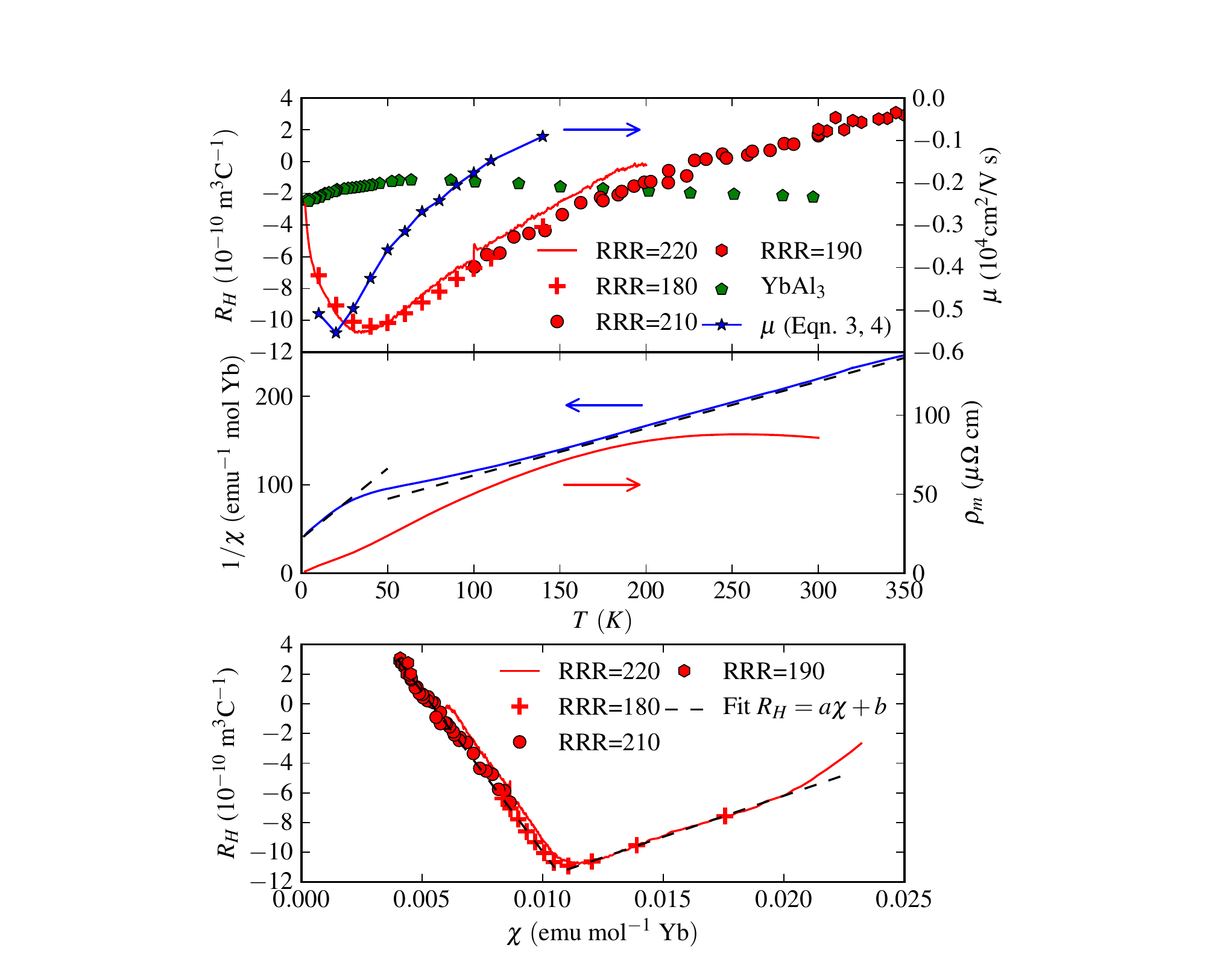}
\put(-210,255){$(a)$}
\put(-210,170){$(b)$}
\put(-210,75){$(c)$}
\caption{$(a)$  Hall coefficient of \bybal~vs. $T$ for different $RRR$ samples measured
	at a field of $\mu_0{}H=0.1$~T // $c$-axis, solid lines and points indicate $T$ and $H$ sweep results, respectively. Also shown are $R_H$ for YbAl$_3$ \cite{cornelius2002two} and the Hall mobility $\mu$ (defined in Eqn. \ref{eqn:2BandModel}, \ref{eqn:2BandParam}) for sample with $RRR=180$.
	$(b)$  Magnetic susceptibility, $\chi \equiv M/H$ (left-hand axis), at $\mu_0{}H=0.1$~T of \bybal~vs. $T$ with Curie-Weiss fits for $T>150$ K and $6<T<15$ K \cite{matsumoto2011anisotropic}. Magnetic resistivity, $\rho_m$ (right-hand axis), obtained by subtracting the estimated lattice contribution to the resistivity from \blbal~\cite{nakatsuji2008superconductivity}.
	$(c)$  Hall coefficient vs. magnetic susceptibility at $\mu_0{}H=0.1$~T.}
    \label{fig:1}
  \end{center}
\end{figure}
 
Fig.~\ref{fig:1}a shows the $T$ dependence of the Hall coefficient ($R_H$) measured by sweeping $T$ or $H$, which agree at low $H$. 
On cooling from 370 K, $R_{H}$ decreases monotonically with positive slope ($dR_{H}/dT > 0$) and changes sign at $T \sim 270$ K.
On further cooling $R_{H}$ forms a clear minimum at 40 K and increases with negative slope at $T<40$~K. 
Furthermore, $R_H$ has a kink just below $T=10$~K corresponding to the Kondo lattice scale ($T^*$), below which HF liquid behavior emerges.
We also include $R_H$ of another mixed valence Yb compound, YbAl$_3$ \cite{cornelius2002two}, for comparison, and the parameter $\mu$ obtained
from the two component fit which we discuss in detail later.

For HF materials including both Ce and Yb compounds, $R_H$ generally peaks at the coherence temperature ($T_{\rm coh}$), below which local moments obtain a band
character \cite{coleman1985theory,fert1987theory,kontani1994theory}. 
Above $T_{\rm coh}$ the AHE due to a lattice of incoherent resonances is known to have the form
$R_A\propto\tilde\chi\rho_{xx}^n$ ($\tilde\chi$ is the scaled susceptibility) with $n$ predicted to be either 1 \cite{fert1987theory} or 0
\cite{coleman1985theory,kontani1994theory}.
Indeed for \bybal, evidence for incoherent scattering is seen by plotting $R_H$ vs. $\chi$ (at $B=0.1$~T
Fig.~\ref{fig:1}c) with $T$ as an implicit parameter. We find that $R_a\propto-\chi\rho_{xx}^0$ ($n=0$) over nearly
one decade of $T$ from 350 K to 50 K, and negative sign as predicted for Yb or $f$-hole materials \cite{kontani1994theory}. 
Therefore, our results indicate that the skew scattering of
local moments persists down to $\sim T_{\rm coh}= 40$ K, where the $4f$ moments become coherent, obtaining a band character.

For most HF materials both $R_H$ and $\rho_{xx}$ peak at approximately the same scale, namely $T_{\rm coh}$, but this is not the case for \bybal.
The magnetic component of $\rho_{xx}$ ($\rho_m$), i.e. with the
lattice contribution subtracted, is
shown in Fig.~\ref{fig:1}b; $\rho_m$ peaks at $\sim$250 K signaling the onset of coherence
in the longitudinal transport, and similar to the fluctuation scale $T_0=200$ K in the specific heat coefficient,
$\gamma\sim1/T_0\ln(T_0/T)$ \cite{nakatsuji2008superconductivity}. 
This is approximately one order of magnitude higher than the (negative) peak in $R_H$ at $\sim40$ K where a small change
in the slope of $\rho_m$ is visible.
Additionally, the magnitude of the $T$ dependence in $R_{\rm H}$ is significantly larger than other Yb based intermediate valence materials,
e.g. YbAl$_3$ with Yb$^{2.71+}$ (Fig.~\ref{fig:1}a) \cite{suga2005kondo}.

Instead, the temperature and magnitude of the peak is similar to those of HF Kondo lattice materials with nearly integral valence such as \yrs~\cite{paschen04hall},
indicating the minimum in $R_H$ is a coherence scale $T_{\rm coh}\sim 40$ K, in addition
to the intermediate valence scale $T_K \sim 250$ K. This additional coherence scale at $\sim40$ K is connected with the emergent Kondo lattice behaviour at low temperatures
\cite{Matsumoto2011Science,matsumoto2011anisotropic}. 
Indeed, instead of Pauli paramagnetism normally seen for strongly intermediate valence systems such as YbAl$_3$ \cite{cornelius2002two}, 
the susceptibility follows the Curie-Weiss (CW) law for local moments
above $\sim 150$ K, and shows a crossover at $T\sim40$~K, before displaying another CW behavior below 15 K \cite{matsumoto2011anisotropic}.
Suggestively, we find $R_H\propto\chi$ even below $T_{\rm coh}\sim40$ K (Fig.~\ref{fig:1}c), which may be related to the possibility of local moment formation as
suggested by the resumption of the CW law. We will discuss this later.

The coefficients obtained by fitting $R_H$ and $\chi$ to Eqn.~\ref{eqn:2} are included in the supplementary material.
In addition, we find that the standard K\"ohler relation for the magnetoresistance is violated in the
range $10<T<110$ K around the minimum
in $\rhall$ (supplementary material). Instead, the modified K\"ohler relation \cite{harris1995violation} holds in this region, evidencing the existence of strong spin or charge fluctuations.

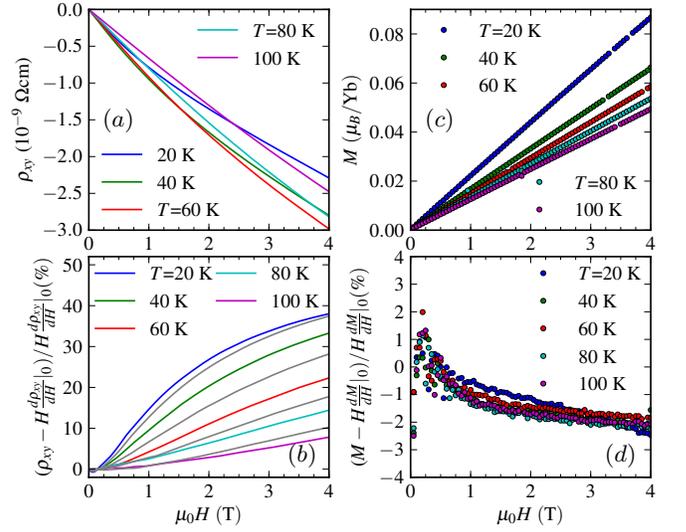
\begin{figure}
\begin{center}
\includegraphics[viewport=65 5 465 331,clip=true,width=\columnwidth]{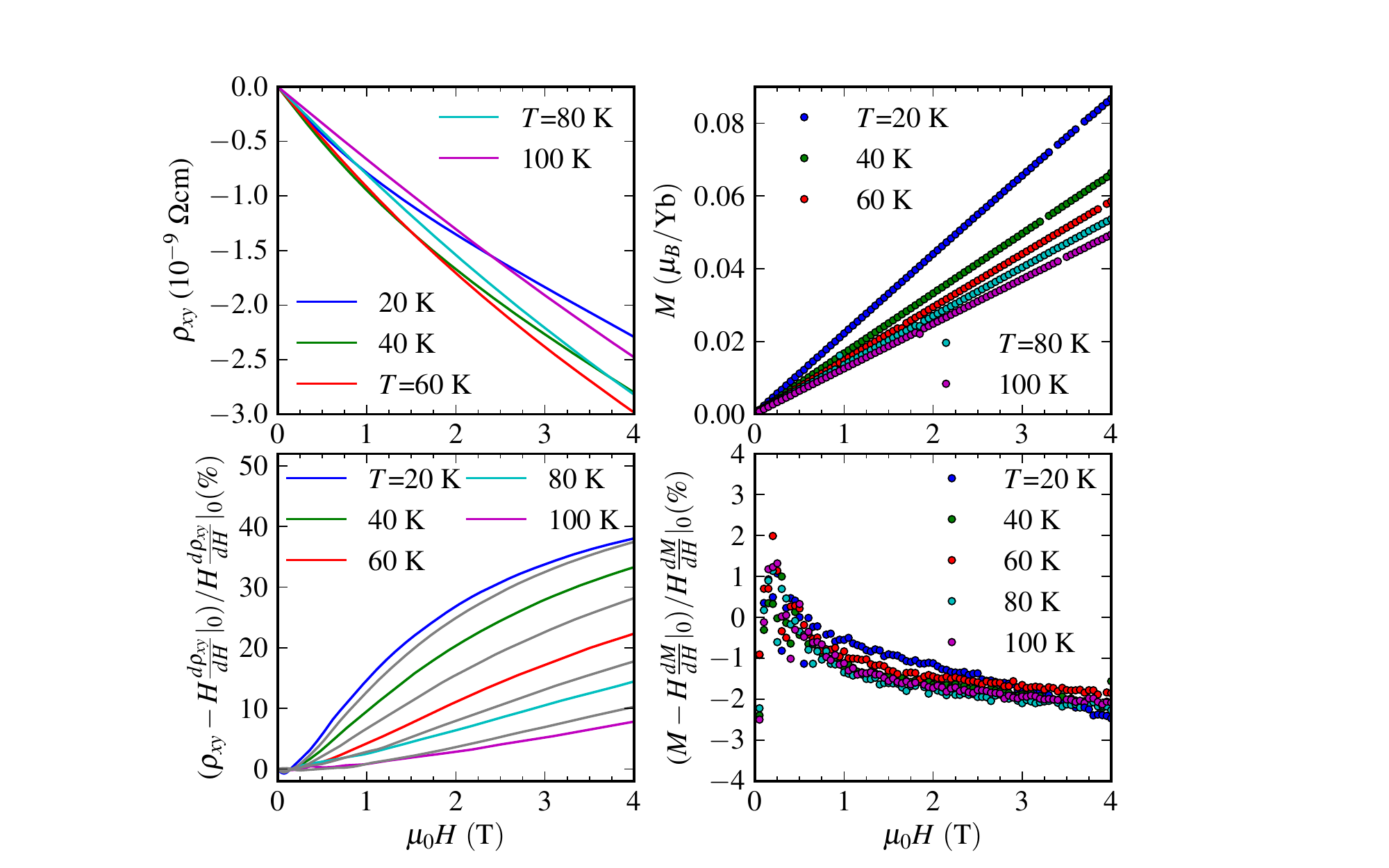}
\put(-210,152){$(a)$}
\put(-140,25){$(b)$}
\put(-88,152){$(c)$}
\put(-20,25){$(d)$}
\caption{$(a)$  Hall resistivity of \bybal~vs. $H\parallel{}c$-axis at several fixed temperatures for a sample with $RRR=180$.
	$(b)$ Percentage deviation of the Hall resistivity from the linear increase with the initial
	slope ($d\rho_{xy}/dH\vert{}_{(H=0)}$) vs. magnetic field with intervening 10 K intervals.
	$(c)$ Magnetization vs. magnetic field measured at
	several fixed temperatures.
	$(d)$ Percentage deviation of the magnetization from the linear increase with the initial slope  ($dM/dH\vert{}_{(H=0)}$).}
    \label{fig:2}
  \end{center}
\end{figure}

Now we discuss the relation between the two distinct Kondo or coherence scales and the evolution of the $H$ dependence of $\rhall$ as a function
of $T$ (Fig.~\ref{fig:2}a). 
$\rhall$ is almost linear in field at $T=100$~K, but becomes increasingly nonlinear at lower temperatures, as is clearly visible in Fig.~\ref{fig:2}b
showing the percentage difference of $\rhall$ from linearity defined using the $H=0$ slope. 
Indeed, the deviation from linearity at $\mu_0{}H=4$~T is $<10\%$ at $T=100$~K, but increases on cooling and levels off to $\sim 40\%$  below $40$~K.
 
Based on the relation $R_H \propto \chi$, one would expect that the AHE increases linearly with $M$ under
field, i.e., $\rhall^{AHE}\propto{}M$. Nonetheless, in sharp contrast to the nonlinearity found in $\rhall$ we find that $M$ has a linear dependence
on the field at the same $T$s. Correspondingly, the percentage
deviation of $M$ from the initial linear slope does not exceed 3 \% at $\mu_0{}H=4$~T even at $T = 20$ K (Fig.~\ref{fig:2}d).
A nonlinear response on $M$ in the AHE has been considered possible in terms of the so-called spin chirality mechanism, wherein the AHE is
instead proportional to the chiral susceptibility which is non-linear in $M$ \cite{machida2009time}. However, the current material has a strong Ising
anisotropy in magnetization, which suppresses the spin chirality and consequently the associated AHE as well.

Given that the non-linearity in the field dependence becomes most pronounced at lower $T$s than the minimum at $T = 40$ K of the Hall coefficient
where coherent behavior is expected, we consider it natural to attribute the nonlinearity in $H$ to the NHE instead, while assuming that the AHE
is linear in $M$ and thus in $H$. 
Such field dependent $R_H$ has been known for the NHE as a consequence of a multi-band effect.
Here we assume the simplest case, namely, 2 independent bands or components; indeed, quantum oscillation (QO) results show that at low $T$ \bybal~has two bands
crossing the Fermi level  \cite{ofarrell-09,tompsett2010role}. Then, analogous to the
Matthiesen rule for longitudinal transport, $\rhall$ may be described by \cite{chambers1952two,arushanov1994hall}:
\begin{equation}
\label{eqn:2BandModel}
\rhall(H)=\frac{R_{H}(0)+R_{H}(\infty)(\mu{}H)^2}{1+(\mu{}H)^{2}}H
\end{equation}
with
\begin{equation}
\label{eqn:2BandParam}
\begin{split}
R_H(0)&=\frac{\rho_1^{2}R_2+\rho_2^{2}R_1}{(\rho_1+\rho_2)^2} & R_H(\infty)&=\frac{R_1+R_2}{R_1R_2}\\
\rho_{xx}&=\frac{\rho_1\rho_2}{\rho_1+\rho_2} & \mu&=\frac{R_1+R_2}{\rho_1+\rho_2},
\end{split}
\end{equation}
where $R_H(0)$ ($R_H(\infty)$) is the value of $R_H$ at $H=0(\infty)$ and $R_i$ and $\rho_i$ are
$R_H$ and $\rho_{xx}$ of each component. The Hall mobility, $\mu\propto\tau/m^*$ (with $\tau$ and $m^*$ the carrier lifetime and
effective mass), controls the field scale of the crossover and when $\mu \rightarrow 0$ the model approaches the single component one i.e. $\rho_i\rightarrow\infty$ and
$\rhall$ becomes linear.

\begin{figure}
\begin{center}
\includegraphics[viewport=11 6 522 267,clip=true,width=\columnwidth]{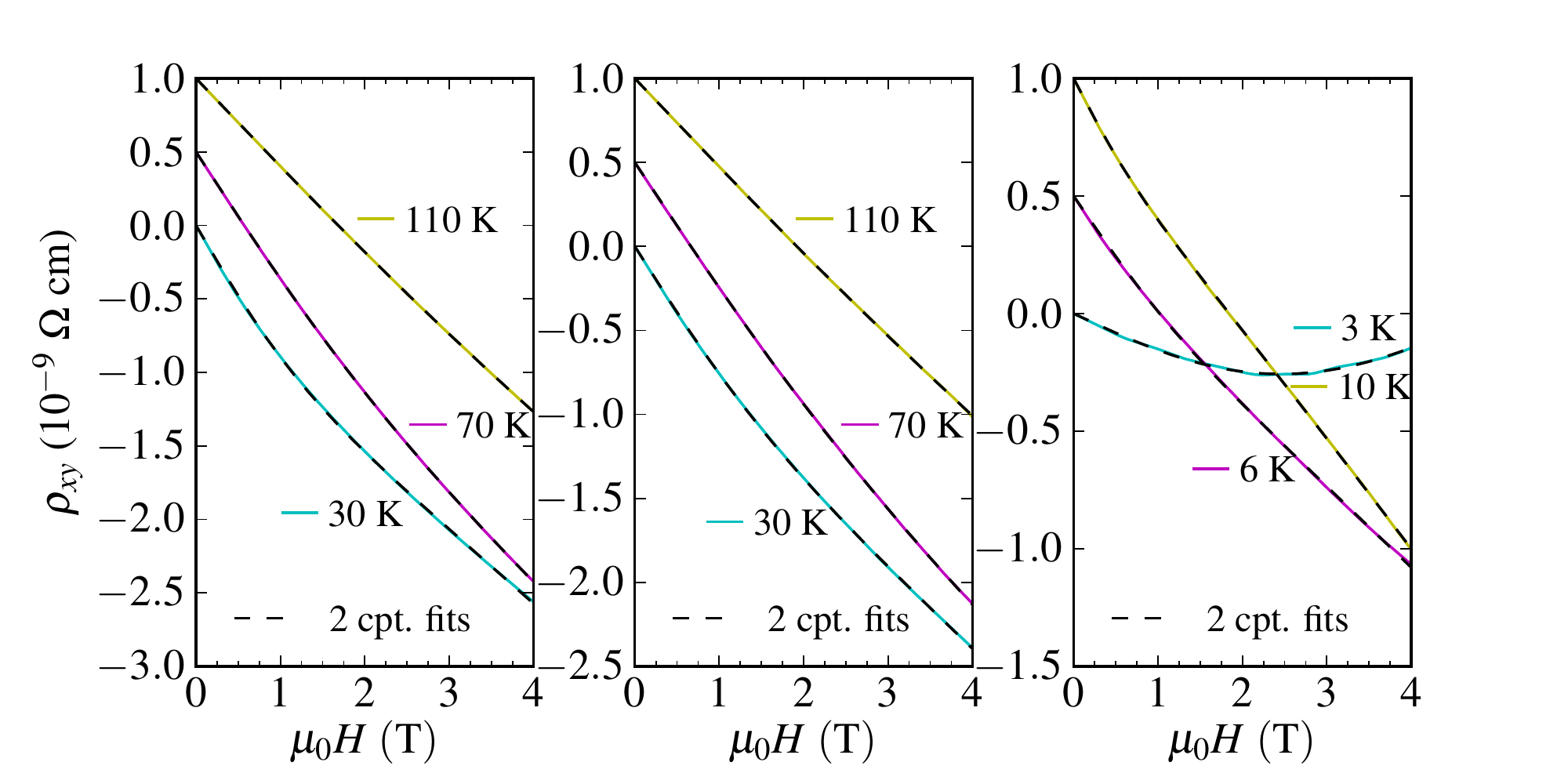}
\put(-95,110){$(b)$}
\put(-170,110){$(a)$}
\put(-20,110){$(c)$}
\caption{$(a)$ Hall resistivity (sample with $RRR=180$) vs. $H$, where successively higher
	$T$s have been offset by +0.5 for clarity, the
	Hall resistivity is fit to the two component model (Eqn.~\ref{eqn:2BandModel}, broken line).
	$(b)$ as in panel $(a)$ for sample with $RRR$=130.
	$(c)$ as in panel $(a)$ for sample with $RRR$=250 measured at lower temperature crossing $T^*$.
	}
    \label{fig:3}
  \end{center}
\end{figure}

\begin{figure}
\begin{center}
\includegraphics[viewport=25 18 535 495,clip=true,width=\columnwidth]{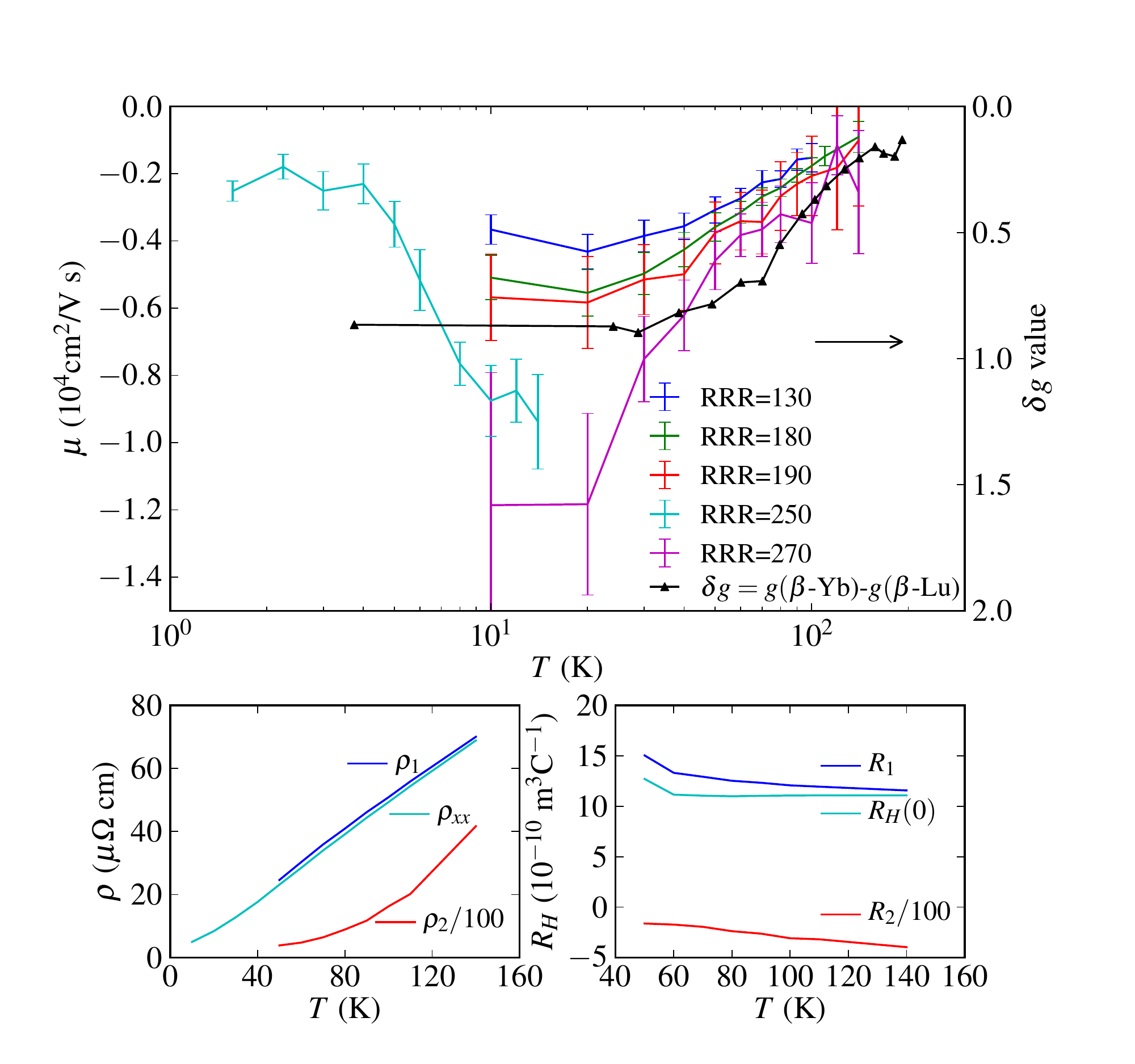}
\put(-215,215){$(a)$}
\put(-215,70){$(b)$}
\put(-35,70){$(c)$}
\caption{$(a)$ Hall mobility $\mu$ (see Eqn.~\ref{eqn:2BandParam}) extracted from the fit to the two component model (see Fig.~\ref{fig:3}b) vs. $T$ for the
	samples in Fig.~\ref{fig:3}a, b, \& c, plotted together with the ESR $g$ value of \bybal \cite{Holanda2011quantum}
	after subtracting the nearly $T$ independent $g$ value for \blbal and consequently multiplied by the constant $C$ for scaling.
	$(b)$ Total measured resistivity together with the fitted resistivities ($\rho_1,~\rho_2$) for the
	two component model vs. $T$. 
	$(c)$ $R_H(0)$ extracted from the two component fit together with the Hall coefficients ($R_{1},~R_{2}$) for each component. For $\rho_2$ and $R_{2}$, the results divided by 100 are shown.
}
    \label{fig:4}
  \end{center}
\end{figure}

As $R_H\propto\chi$ and $M \propto H$, we assume that at fixed temperature $\rhall^{AHE}=R_{a}\chi(\mu_{0}H=0.1 {\rm T})H$. Notably for our analysis, subtracting a linear
in $H$ term from
Eqn.~\ref{eqn:2BandModel} does not affect the parameter $\mu$ because $\mu$ does not appear in the coefficient of the term linear in $H$.
Therefore, to determine the $T$ dependence of the mobility $\mu(T)$ (Fig.~\ref{fig:4}a and Fig.~\ref{fig:1}a), we
may use $\rhall$, without any AHE subtraction, for fitting to the two-component model.
Significantly, the fitting successfully reproduces all the field-dependent data of $\rhall$
obtained over a wide temperature range covering 2 K $<T<150$ K, and irrespective of crystal quality ranging from $RRR=130$, 180 to 250 (Fig.~\ref{fig:3}a, b, and c). 
According to Eqn.~\ref{eqn:2BandModel}, the increase in $\vert\mu(T)\vert$ from zero itself corresponds to the emergence of the $2^{nd}$ component
on cooling in the  model. 
As we will discuss, the analysis suggests emergent \emph{coherent} Hall transport of the $2^{nd}$ component below the characteristic scale of $T_{\rm coh} = 40$ K where $\vert\mu\vert$ is largest as is clearly visible in Fig.~\ref{fig:1}a.

The overall $T$ dependence for $\vert\mu\vert$ is found to be the same for all the samples. Namely, at high $T$s, $\mu$ is vanishingly small, 
corresponding to the field linear response of $\rhall(B)$. However, below 100 K $\vert\mu\vert$ increases on cooling and finally levels off below
40 K, signaling the formation of the coherent transport, consistent with the coherence peak found in the $R_H$ at $T_{\rm coh} = 40$ K.
Finally, in the quantum critical region below $T^*$ $\vert\mu\vert$ gradually decreases, most likely due to the increase in $m^*$ as the result of QC fluctuations.

The larger $\vert\mu\vert$ was confirmed for the sample with larger $RRR$ at low temperatures below $\sim 40$ K, while  no $RRR$ dependence in $\mu$ was observed at high
temperatures, suggesting that strong magnetic or charge fluctuations determine the mean free path and thereby the mobility.
The stronger dependence of $\mu$ on $RRR$ or static disorder at the lower $T$ below  $\sim 40$ K suggests again the emergent coherence.

Another evidence of the coherence below $\sim 40$ K is the scaling between the $T$ dependence of $\mu(T)$ and the Land\'e $g$ value of \bybal~obtained by
electron spin resonance (ESR) after subtracting that of \blbal~(Fig.~\ref{fig:4}a) \cite{Holanda2011quantum}. Given that the behavior is due to the $conduction$
ESR, as inferred from the $T$ independent behavior of the integrated intensity,
the increase of the $g$ value on cooling indicates that the conduction electron acquires $f$-electron characters following the development of the
Kondo resonance below $T_{\rm coh} = 40$ K.
The scaling behavior between  $\mu(T)$ and $g(T)$ thus suggests that the $2^{nd}$ component acquires the coherence below $T_{\rm coh}$.

To extract the other individual parameters ($R_1$, $R_2$, $\rho_1$, $\rho_2$) of Eqn.~\ref{eqn:2BandParam}, we must subtract the AHE from $R_H$ so that only the NHE is fitted.
We observed that $R_H\propto\chi$ in two regimes $T>60$~K and $T<40$~K but with a dramatic change in coefficients indicating a significant change in electronic structure following the formation of coherence.
We attribute the AHE at
$T>60$~K to resonant skew scattering and, on the basis of these results, we believe the AHE can only be unambiguously subtracted at $T>50$ K,
as described in the supplementary material. Following the subtraction, $R_H$ becomes positive (holelike). We find $R_2$ and $\rho_2$ are approximately
2 orders of magnitude larger than the values for the $i=1$ component and that $R_2$ is electronlike (Fig. \ref{fig:4}b \& c). This confirms that the 2nd component dominates the total 
mobility $\mu\approx\mu_2$. Using the Drude model $R_i\propto1/n_i$, a straightforward interpretation of these results is that the carrier density $n_2$ is
$\sim 10$ times smaller than $n_1$ at 50 K and gradually increases as $T$ is lowered. 


The phenomenological two component model thus naturally accounts for two different coherent energy scales that
respectively governs the longitudinal transport and the Hall mobility. 
Namely, the above analysis indicates that  the $i=1$ ($1^{\rm st}$) component is much more conductive
than the $i=2$ ($2^{\rm nd}$) component and the coherence $T_{\rm K} \sim 250$ K is the scale set by the $1^{\rm st}$ component. 
Furthermore, the increase in $\mu$ confirms that it is the $2^{\rm nd}$ component that
has coherence temperature of $\sim 40$ K, one order of magnitude smaller than $T_{\rm K}\sim250$ K. 
Thus, it is the $2^{\rm nd}$ component that should be the origin of the dichotomy, namely, the emergent Kondo lattice and quantum critical behaviors in the strongly intermediate valence state.
Naturally this result asks the question of how the interaction with the $f$ electron can split the conduction electron into 2 essentially independent components, one of which remains
partially incoherent even below it's dominant coherent $T$ scale.

The hint may be hidden in the electronic structure set by the crystal symmetry. 
The lattice of Yb ions in \bybal~can be viewed as a layered distorted honeycomb structure, in which each layer containing Yb and Al is separated by B sheets \cite{macaluso2007crystal,nevidomskyy2009layered}. 
Both local density approximation (LDA) and tight binding calculations emphasized the importance of the hybridization between Yb and the adjacent boron layers and the anisotropic
hybridization that results from this layered structure \cite{ofarrell-09,tompsett2010role};
indeed, on the basis of the crystal electric field analysis as well as an exceptionally large transport anisotropy in \aybal~($\rho_{ab}/\rho_c\approx15$) it was suggested that there may be a node in the hybridization
parallel to the $c$-axis and at $k_x{}=k_y{}=0$ \cite{matsumoto2011anisotropic,ramires2012beta}. 

Therefore, we speculate that the origin of two components
of the Hall effect is the Yb ion hybridizing differently with distinct conduction electrons which are separated in momentum space and do not interact except via the $f$ electrons.
In this case, the component associated with the node in the hybridization must be $2^{\rm nd}$ component that has a low mobility and $T_{\rm K}\sim 40$ K.
The apparent incoherent skew scattering observed below $T_{\rm K}$ suggests the role of the node which in principle could be the
source of the incoherent scattering from the unquenched local moment down to zero temperature, and therefore might lead to the quantum criticality. 

Taking the possibility of a node further, in the context of our results, we consider the QO FS; two bands were found to cross the Fermi level,
the holelike band 89 and the electronlike band 91 \cite{ofarrell-09,nevidomskyy2009layered,tompsett2010role}. The sign of $R_H$ suggests that Band 89 and 91 correspond to the $1^{st}$ and  $2^{nd}$ components, respectively.
Indeed, Band 91 has a quasi-2D portion centered around $k_x=k_y=0$ where the small hybridization and low $T_{\rm K}$ are expected. However, Band 89 is squashed along $k_z$ with portions that have a large $k_x$ and $k_y$ leading to larger hybridization
on this FS and a higher $T_{\rm K}$. Thus, our results are consistent with both with the experimental FSs and the possible node in the hybridization.

A recent theory \cite{ramires2012beta} suggests the FS portion nearby the hybridization node is responsible for the singular behavior at zero field and ambient pressure.
This is apparently characterized by a low carrier concentration ($\sim 10$ \%) of the total and localized moment behavior down to low temperatures. Our results find
the carrier concentration of the $2^{\rm nd}$ component is gradually increasing on cooling and $\sim 9.5$ \% at 50 K.
This as well as the apparent residual incoherent AHE and the correspondence with the experimental FS suggest the $2^{\rm nd}$ component be consistent with
the nodal portion in the phenomenological theoretical model. Further low $T$ investigation of the Hall effect is clearly warranted to clarify how this component exhibits the unconventional QC.
  
\begin{acknowledgments}
We acknowledge K. Ueda,  P. Coleman, K. Miyake and S. Onoda for useful discussions and K. Kuga for experimental assistance.
This work was partially supported by Grants-in-Aid (No.21684019) from JSPS, by Grants-in-Aids for Scientific Research on Innovative Areas ``Heavy Electrons" of MEXT, Japan and Toray Science Foundation. E O'F acknowledges a Overseas Postdoctoral Fellowship from JSPS.
\end{acknowledgments}

\cleardoublepage
\onecolumngrid
\appendix

\renewcommand{\thefigure}{S\arabic{figure}}
\renewcommand{\thetable}{S\arabic{table}}
\renewcommand{\theequation}{S\arabic{equation}}

\renewcommand{\bibnumfmt}[1]{[S#1]}
\renewcommand{\citenumfont}[1]{\textit{S#1}}

\setcounter{figure}{0}
\setcounter{equation}{0}

\section{Supplementary Online Material for:\\Evolution of $c$-$f$ hybridization and a two component Hall effect in \bybal}
 

 
 
 
 

\section{Methods}

Single crystals were grown using Al flux method \cite{macaluso2007crystal}. 
We performed Hall effect measurements of several single crystals of \bybal~with different residual resistivity ratios ($RRR$). 
The Hall voltage and longitudinal resistivity were measured simultaneously using a 5-wire method with current applied within the $ab$-plane and magnetic field parallel to the $c$-axis.

$\rhall$ was obtained from the antisymmetric part of the transverse voltage. 
$R_H$ is defined as $R_H=\rhall/H$ and $d\rhall/dH$, respectively for $T$ and $H$ sweep measurements. At low fields when $\rhall$ is linear in \bybal~these definitions agree.
 

\section{Coefficients of the Anomalous Hall effect}

The Hall coefficient is described as the sum of normal and anomalous parts, according to \cite{nagaosa2010anomalous}:
\begin{align}
\label{eqn:S1}
\rhall(T,H)&=R_{N}(T,H) H + R_{a}(T,H) M(T,H)\\
\label{eqn:S2}
R_H&=R_N+R_a\chi\equiv R_N+R_A.
\end{align}
where $T$, $H$,  $M$ and $\chi \equiv M/H$ are temperature, applied magnetic field, magnetization, and magnetic susceptibility, respectively. 
We note that the separation of the NHE coefficient $R_N$ and AHE coefficient $R_A\equiv R_a\chi$ in Eqn.~\ref{eqn:2} is frequently ambiguous,  requiring careful consideration of the underlying mechanism of the AHE and  the electronic structure.

In Fig. 1C of the main article we observe two regions where $R_H\propto\chi$ corresponding to two temperature ranges. The coefficients of Eqn.~\ref{eqn:2} are as in Table
~\ref{tab:AHE}.
\begin{table}[H]

\begin{center}
\caption{Coefficients of Eqn. \ref{eqn:S2} for temperature ranges above and below the minimum in $R_H$}
\begin{tabular}{lll}
\label{tab:AHE}
Temperature range $(K)$ & $R_N~(10^{-10}m^{3}C^{-1})$ & $R_a~(10^{-10}m^{3}C^{-1}\rm{emu}^{-1})$ \\
\hline
$10\leq{}T\leq40$ & 11.6 & -2150\\
$60\leq{}T\leq300$ & -17.3 & 550\\
\hline\end{tabular}
\end{center}
\end{table}

\begin{figure}[!ht]
\begin{center}
\includegraphics[viewport=37 24 507 525,clip=true,width=0.6\columnwidth]{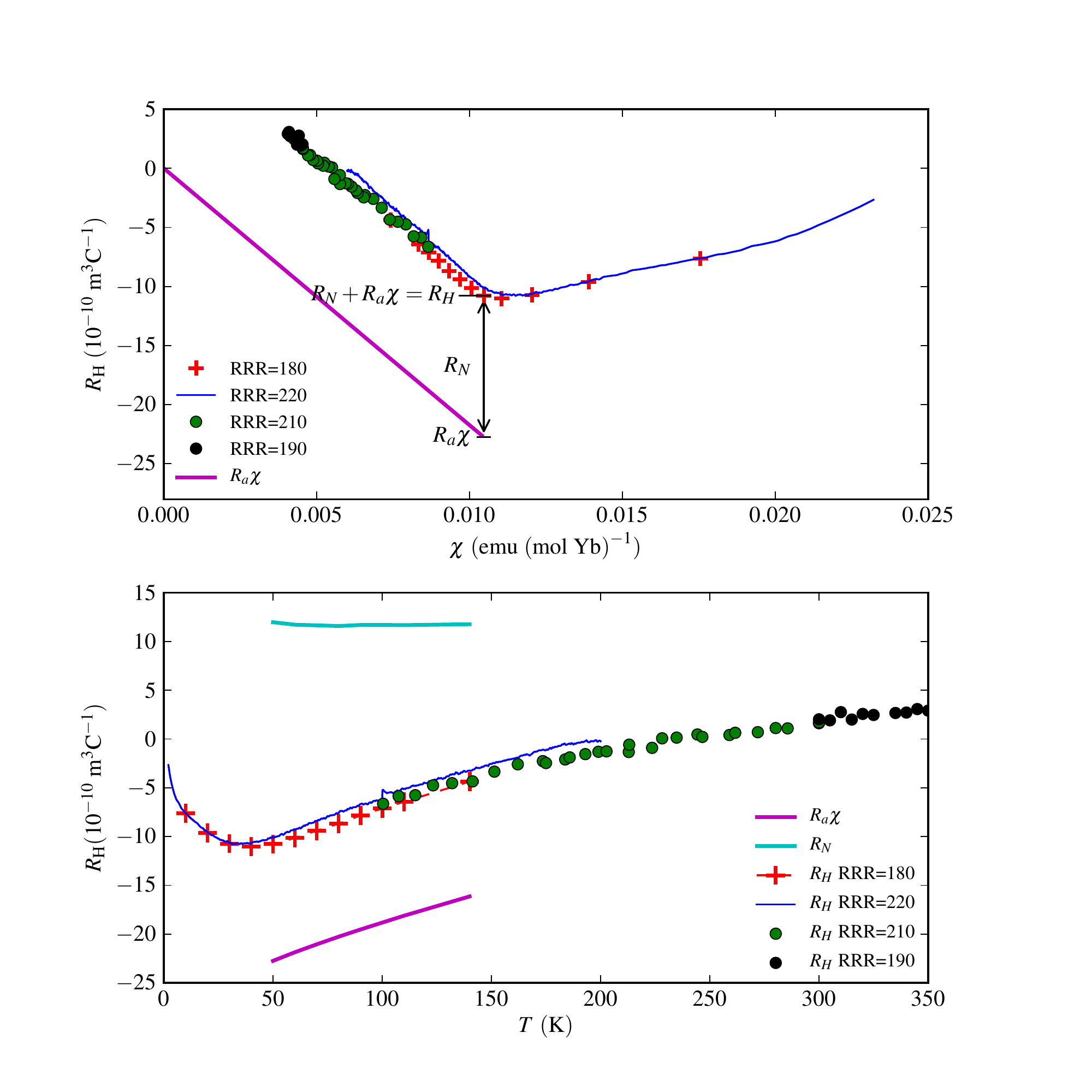}
\put(-30,310){$(a)$}
\put(-30,140){$(b)$}
\caption{$(a)$: Procedure used for subtracting the anomalous Hall effect (AHE) from the total Hall effect.
	$(b)$: Proposed temperature dependence of the normal and anomalous Hall coefficients, $R_N$ and $R_A$ in the temperature
	range where the subtraction was performed.
	}
    \label{fig:S1}
  \end{center}
\end{figure}

We use these coefficients to subtract the anomalous Hall effect (AHE) part from the total Hall coefficient, as described in the main text. The dependence of both total Hall effect and AHE on the magnetic susceptibilty is
illustrated in Fig.~\ref{fig:S1}A and the dependence of both the AHE and the normal Hall effect (NHE) coefficients on
temperature is illustrated in Fig.~\ref{fig:S1}B.
 
\section{Curie-Weiss fits for magnetic susceptibility}

The analysis of the magnetic suceptibility of \bybal~is described in detail in \cite{matsumoto2011anisotropic}, here we reproduce the results of that analysis. The Curie-Weiss relation for fields along the $c$-axis in an Ising system is
\begin{equation}
\label{eqn:CW}
\chi_c=\frac{C}{T+\Theta_{\rm W}}
\end{equation}
where \Weiss~is the Weiss temperature and $C=N_AI_z^2/k_B$ with $N_A$ and $k_B$ the Avagadro and Boltzmann constants and $I_z$ the Ising moment.

The constants for \bybal at high and low temperatures above and below the minimum in $R_H$ are as in Table
~\ref{tab:chi}.
\begin{table}[H]
\begin{center}
\caption{Coefficients of Eqn. \ref{eqn:CW}}
\begin{tabular}{lll}
\label{tab:chi}
Temperature range (K) & \Weiss~$(K)$ & $I_z$ ($\mu_B$) \\
\hline
$150\leq{}T\leq350$ & 108 & 2.24\\
$60\leq{}T\leq300$ & 25 & 1.3\\
\hline\end{tabular}
\end{center}
\end{table}

\section{K\"ohler's relation for magnetoresistance}

\begin{figure}
\begin{center}
\includegraphics[viewport=2 3 395 265,clip=true,width=0.7\columnwidth]{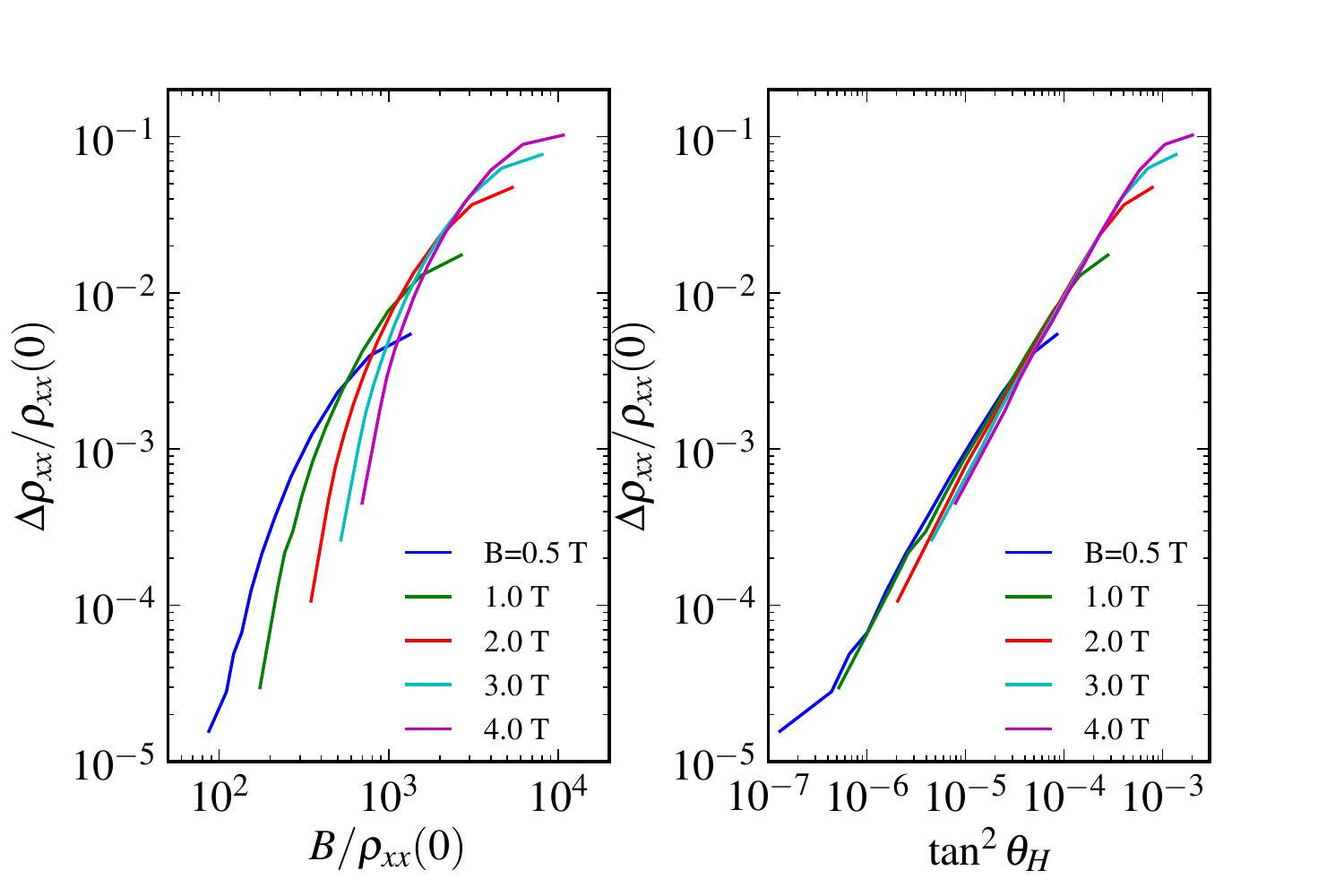}
\put(-300,220){$(a)$}
\put(-120,220){$(b)$}
\caption{$(a)$: K\"ohler magnetoresistance plot ($\Delta\rho_{xx}=\rho_{xx}(B)-\rho_{xx}(0)$) in the temperature range of $10\leq{}T\leq110$ K. If the Kohler rule is valid, all the data should collapse on top.
	$(b)$: Modified K\"ohler plot  using the square of $\tan\theta_H=\rho_{xy}/\rho_{xx}$ for the horizontal axis. All the data reasonably overlap each other, indicating the modified version holds for \bybal.
}
    \label{fig:SOM-Kohler}
  \end{center}
\end{figure}

The K\"ohler relation for the magnetoresistance is usualy expressed as
\begin{equation}
\frac{\rho_{xx}(B)-\rho_{xx}(B=0)}{\rho_{xx}}=\frac{\Delta\rho_{xx}}{\rho_{xx}}=F\Big(\frac{B}{\rho_{xx}}\Big)
\end{equation}
where $F$ is a function that depends on the electronic structure. This is expected to hold in conventional metals where a single scattering time determines the transport
properties. Therefore, a so called K\"ohler plot of $\Delta\rho_{xx}/\rho_{xx}$ vs. $B/\rho_{xx}$ on a logarithmic scale is expected to collapse the data from all fields
on a single line. Fig.~\ref{fig:SOM-Kohler} shows the K\"ohler plot for \bybal{},
indicating that the K\"ohler relation does not hold in this material.

Several materials have been reported for which the conventional K\"ohler scaling does not hold. Broadly speaking these belong to the class of strongly correlated
electron materials wherein strong spin or charge fluctuations influence the transport. Prominent examples are the cuprate high temperature superconductors \cite{harris1995violation} and
$f$-electron based heavy electron materials \cite{nakajima2007non}. An alternative, so called modified K\"ohler, scaling was proposed for the cuprates \cite{harris1995violation}
\begin{equation}
\frac{\Delta\rho_{xx}}{\rho_{xx}}\propto\tan^2\theta_H,
\end{equation}
where $\theta_H$ is the Hall angle ($\tan\theta_H=\rho_{xy}/\rho_{xx}$). Figure~\ref{fig:SOM-Kohler} shows this scaling relation for \bybal  using raw data for $\rho_{xy}$ (\emph{i.e.} without subtracting the AHE), indicating that it holds
reasonably accurately in the temperature range $10<T<110$ K.

The modified K\"ohler rule was proposed to consider the presence of an additional lifetime $\tau_H$ that governs the transverse transport and comes from antiferromagnetic
fluctuations. For the case of \bybal, the failure of the K\"ohler rule and the success of the modified K\"ohler relation down to 10 K suggests that strong spin and charge
fluctuations affect the transport despite the onset of coherence at $T_{\rm K} = 250$ K.

\end{document}